\documentstyle[12pt]{article}
\input epsf
\newcommand{\be}[1]{\begin{equation} \label{(#1)}}
\newcommand{\ee}{\end{equation}}
\newcommand{\ba}[1]{\begin{eqnarray} \label{(#1)}}
\newcommand{\ea}{\end{eqnarray}}
\newcommand{\nn}{\nonumber}
\newcommand{\rf}[1]{(\ref{(#1)})}
\def\21{$SU(2) \otimes U(1) $}
\def\nt{\hbox{$\nu_\tau$ }}
\def\rp{$R_p \hspace{-1em}/\;\:$ }

\def\lv{$L \hspace{-0.6em}/\;\:$ }
\def\rpm{R_p \hspace{-0.8em}/\;\:}

\def \znbb {\beta\beta_{0\nu}}

\def \emass {\langle m_{\nu} \rangle}

\begin{document}
\hfill hep-ph/yymmdd\\
\hspace*{1cm}\hfill FTUV/98-100\\
\hspace*{1cm}\hfill IFIC/98-101\\
\bigskip
\begin{center}
{\Large \bf Neutrinoless double beta decay in Supersymmetry with
bilinear R-parity breaking}

\bigskip
{\large M. Hirsch\footnotemark[1] and J.~W.~F. Valle\footnotemark[2]}

\bigskip
{\it Instituto de F\'{\i}sica Corpuscular -- C.S.I.C. \\ Departamento
de F\'{\i}sica Te\`orica, Universitat of Val\`encia \\ 46100
Burjassot, Val\`encia, Spain\\ http://flamenco.uv.es//}

\end{center}


\begin{abstract}

We reanalyze the contributions to neutrinoless double beta ($\znbb$)
decay from supersymmetry with explicit breaking of R-parity. Although
we keep both bilinear and trilinear terms, our emphasis is put on
bilinear R-parity breaking terms, because these mimic more closely the
models where the breaking of R-parity is spontaneous.  Comparing the
relevant Feynman diagrams we conclude that the usual mass mechanism of
double beta decay is the dominant one. From the non-observation of
$\znbb$ decay we set limits on the bilinear R-parity breaking
parameters of typically a (few) 100 $keV$.  Despite such stringent
bounds, we stress that the magnitude of R-parity violating phenomena
that can be expected at accelerator experiments can be very large,
since they involve mainly the third generation, while $\znbb$ decay
constrains only the first generation couplings.
\end{abstract}
\bigskip
\bigskip
\footnotetext[1]{mahirsch@neutrinos.uv.es}
\footnotetext[2]{valle@neutrinos.uv.es}

\section{Introduction}

The most well-studied realization of supersymmetry has a conserved
R-parity. This is a multiplicative quantum number which can be defined
as $R_P = (-1)^{3B+L+2S}$, where $B$ and $L$ are the baryon and lepton
number and $S$ the spin of the corresponding particle. This defining
property of the Minimal Supersymmetric Standard Model (MSSM)
\cite{HK85}  is theoretically {\sl ad hoc} since the origin of 
R-parity conservation is unknown. Since the early days it was
recognized that the violation of R-parity could arise as residual
effects in larger theories~\cite{HS84} or spontaneously due to a
sneutrino vacuum expectation value (VEV)~\cite{oldRPV}.  We find the
models with pure spontaneous breaking more well-motivated than
explicit breaking models, in the sense that they put R-parity breaking
on a dynamical footing. They have only lepton number violating (\lv)
terms and are therefore in agreement with proton stability.  The
simplest models of this type (with the \21 gauge structure) are
characterized by the existence of a physical Goldstone boson,
generically called majoron. In the early versions the majoron was an
isodoublet (the imaginary part of the sneutrino) ~\cite{oldRPV}; these
are now ruled out by LEP measurements of the invisible Z
width~\cite{oldsponrpv}.  However, in the modern variants of
supersymmetric models with spontaneous R-parity breaking the majoron
is a singlet (such as a right-handed sneutrino) and therefore can not
be produced in the decays of the Z. Such (\rp) models are both
theoretically attractive and phenomenologically viable
\cite{MV90,val98}. However, quite generally study of broken R-parity 
phenomenology has received a lot of attention recently
\cite{ConchaV,finnish,RPVothers,giudice}.

Especially interesting in view of the mounting evidence for non-zero
neutrino masses \cite{SuperK,Solarnu,status98} is the fact that \rp
models allow for finite - and in principle calculable - neutrino
masses, and moreover could even explain their observed smallness
without invoking some high-energy scale physics \cite{RPVmnu}, like
one does for example in the see-saw mechanism \cite{seesaw}.

Neutrinoless double beta ($\znbb$) decay is a $\Delta L=2$ process and
therefore one naturally expects it to occur in models with lepton number
violation in the Lagrangian. Even though there is a variety of
mechanisms inducing $\znbb$ decay in gauge theories, one can show
that whatever the leading mechanism is at least one of the neutrinos
will be a Majorana particle~\cite{SV82}, as illustrated in the
black-box diagram of Fig.~1.  This well-known argument establishes a
deep connection between Majorana neutrino masses and $\znbb$ decay: in
gauge theories one can not occur without the other being present. The
same remains true in supersymmetric theories \cite{theorem}, where
moreover one can show that also the supersymmetric partner of the
neutrino must have a $B-L$ violating {\sl Majorana-like} mass term, if
a Majorana mass of the neutrino exists \cite{theorem}.  Turning the
argument around, one expects that the observed absence of $\znbb$
decay allows to derive stringent limits on \rp parameters. This has
been shown for models with explicit trilinear R-parity breaking in
\cite{hir9598,fae97}.  In the current work we will concentrate on the
bilinear \rp terms.

$\znbb$ decay within supersymmetric models with bilinear R-parity 
breaking has been considered previously in \cite{FKS98}. In our 
derivation the final expression of the decay rate (and the 
corresponding nuclear matrix elements), however, differ from 
the one given in \cite{FKS98}, details are given in section 3. 

In an exactly supersymmetric and $SU(2)_L$-invariant world R-parity
Violation (RPV) can be moved around the Lagrangian by field
redefinitions \cite{HS84}.  This observation lead to the claim, that
the bilinear RPV terms are unphysical and should be discarded. However
it has been shown that in realistic softly broken supersymmetric
models it is in general impossible to eliminate the bilinear terms
\cite{epsrad}. Moreover, as already stated, it is possible that
R-parity is a symmetry (maybe approximate) at the Lagrangian level,
broken only by the ground state~\cite{MV90}.  Such scenarios lead only
to \lv being therefore consistent with baryon number conservation.
Models with bilinear terms can be imagined as a truncated version of
the full dynamical models \cite{epsrad} and as such offer a systematic
way to investigate RPV at the phenomenological level.

This paper is organized as follows. In the next section we introduce
our conventions and notations. Section 3 derives the effective
Lagrangian of $\znbb$ decay in \rp SUSY with bilinear terms, while
section 4 summarizes the necessary nuclear physics input. Section 5
then analyzes the constraints $\znbb$ decay can put on bilinear \rp
parameters.

\section{Supersymmetry with explicit R-parity breaking terms}

In this section, we set up our definitions and notation. Our
conventions are usually such that in the limit of vanishing \rp
parameters the MSSM notation of \cite{HK85} is recovered.

In the presence of bilinear R-parity and lepton number violation there
is no distinction between the lepton doublet and the Higgs doublet
superfield giving mass to the down-type quarks. This fact can be
accounted for by defining a superfield ${\hat \Phi}$ as

\be{defphi}
{\hat \Phi} = ({\hat H_1},{\hat L_1},{\hat L_2},{\hat L_3}).
\ee
For the MSSM field content the most general gauge invariant 
form of the renormalizable superpotential can then be written 
as 

\be{superpotphi}
W = \epsilon_{ab} \Big[ 
    \lambda_{e}^{IJk} {\hat \Phi}_I^a {\hat \Phi}_J^b {\hat E}_k^C
  + \lambda_{d}^{Ijk} {\hat \Phi}_I^a {\hat Q}_j^b {\hat D}_k^C
  +  h_{u}^{jk} {\hat Q}_j^a {\hat H}_2^b {\hat U}_k^C
  +  \mu^I {\hat \Phi}_I^a {\hat H}_2^b \Big] .
\ee
Here, ${\hat Q}$ and ${\hat D}^C$, ${\hat U}^C$ are the quark doublet
and singlets superfields, respectively, ${\hat E}^C$ is the lepton
singlet superfield and ${\hat H}_2$ the Higgs superfields with
$Y({\hat H}_2) = 1$ responsible for the up-type quark masses, with
$h_{u}^{jk}$ being the corresponding Yukawa couplings.  The indices
$j,k~=~1,2,3$ denote generations, whereas $I,J~=~0,1,2,3$. The indices
$a,b$ are $SU(2)$ indices.  In the basis \rf{defphi} one can separate
$W$ into an R-parity conserving and an R-parity violating (\rp) part
\be{superpot}
W = W_{R_P} + W_{\rpm},
\ee
where the $R_P$ part is simply given by taking $I$ or $J$ to be 
zero and $i,j$ respectively. Then $\lambda_{e}^{0jk}= \frac{1}{2}
h_{u}^{jk}$ and $\lambda_{d}^{0jk}=h_{d}^{jk}$. The \rp part 
contains the remaining components, 
\be{WRPV} 
W_{\rpm} = \lambda_{e}^{ijk}{\hat L}_i{\hat L}_j {\hat E}^C_k 
         + \lambda_{d}^{ijk}{\hat L}_i {\hat Q}_j {\hat D}^C_k 
         + \epsilon_i {\hat L}_i {\hat H}_2 
\ee
where $\mu_i \rightarrow \epsilon_i$, to make contact with the
notation of \cite{epsrad}. If $W_{\rpm}$ would be the only source of
RPV one could easily rotate the field ${\hat \Phi}$ into a basis
${\hat \Phi}'$ with $\mu^{I'}=(\mu',0,0,0)$ effectively eliminating
the bilinear terms. However, another source of RPV is found in the
soft supersymmetry breaking part of the scalar potential. It contains
the terms:
\ba{vsoftrp} \nn
V^{soft}_{\rpm} & = & 
{\tilde A}_{e}^{IJk}{\tilde{\Phi}}_I {\tilde {\Phi}}_J {\tilde E^C}_k + 
{\tilde A}_{d}^{Ijk}{\tilde {\Phi}}_I {\tilde Q}_j {\tilde D^C}_k 
+ B^I {\tilde {\Phi}}_I  H_2 \\
&+& (m_{IJ}^2 + \mu_I\mu_J) {\tilde \Phi}_I {\tilde \Phi}_J^{\dagger} 
+ \cdots + h.c. ,
\ea
where the dots represent terms not interesting for the discussion 
here. Rotating the superpotential as discussed above, it is easy 
to see that as long as the $B^I$ are not exactly parallel to the 
$\mu^I$ the effects of the bilinear terms of the superpotential 
in the rotated basis will reappear in the soft SUSY breaking terms.

Let us briefly discuss the appearance of bilinear terms $B^I {\Phi}_I
H_2$. The presence of these terms imply that the tadpole equations for
the sneutrinos are non-trivial, i.e. the VEVs of the sneutrino fields
in general are non-zero \cite{FiniteVEVs}. As a result the MSSM
vertices ${\tilde Z}\nu{\tilde \nu}$ and ${\tilde W}e{\tilde e}$
create gaugino-lepton mixing mass terms ${\tilde Z}\nu<{\tilde \nu}>$
and ${\tilde W}e<{\tilde \nu}>$. This mixing then leads in the
physical mass eigenstate basis to lepton number violating (LNV)
interactions. The LNV vertices relevant to $\znbb$ decay will be
discussed in the next section. The mass eigenstate fields are defined
from the weak eigenstates by means of an orthogonal transformation,
\be{wm0}
\Psi_{0,i} = \Xi_{ij}\Psi'_{0,j},
\ee
\be{wmp}
\Psi_{(\pm)i} = \Delta^{\pm}_{ij}\Psi'_{(\pm)j},
\ee
with the weak basis states in two component notation defined as 

\be{ntrlwk}
{\Psi'_{0}}^T = (\psi_{L_1}^1, \psi_{L_2}^1, \psi_{L_3}^1, 
                 -i\lambda', -i \lambda_3, 
                 \psi_{H_1}^1,\psi_{H_2}^2),
\ee
\be{chrgmns}
 \Psi_{(-)}'^T = (\psi^{2}_{L_1},\psi^{2}_{L_2},\psi^{2}_{L_3},
                 -i\lambda_{-},\psi_{H_1}^{2}),
\ee
\be{chrgpls}
 \Psi_{(+)}'^T = (\psi_{R_1}^c,\psi_{R_2}^c,\psi_{R_3}^c,-i\lambda_{+},
                 \psi_{H_2}^{1}),
\ee
where $\psi_{L_i}^1$ ($\psi_{L_i}^2$) are the neutrino (charged
lepton) fields, $-i\lambda'$ and $-i \lambda_3, \lambda_{-}$ are the
unmixed photino and gaugino states, respectively, and
$\psi_{H_{1,2}}^{1,2}$ are the neutral components of the Higgs
doublets, and the Higgs and lepton fields are written in the original
basis $\Phi$.

The mixing matrices $\Xi$ and $\Delta^{\pm}$ diagonalize the
neutrino-neutralino and charged lepton-chargino mass matrices.
Details are given in the appendix. We note in passing that the neutral
mass matrix has such a texture that only one neutrino gains a
tree-level mass, even if all of the $\epsilon_i$ are non-zero. The
degeneracy of the remaining two neutrinos, massless at tree-level,
is however lifted at the 1-loop level \cite{1-loop}.

\section{Effective Lagrangian of $\znbb$ decay}

Here we derive the effective Lagrangian describing $\znbb$ decay in
the RPV model.  Formally one can define the amplitude for $\znbb$
decay as,
\be{defr0n}
R_{\znbb} = \langle (A,Z+2),2e^{-}|
            {\cal T}\Big[exp\Big(i\int d^4x {\cal L}_{eff}(x)\Big)\Big]
            |(Z,A) \rangle.
\ee
For the derivation of ${\cal L}_{eff}$ it is convenient to decompose
the different \rp contributions as shown in figure 2.  As indicated in
the figure, in the \rp MSSM the effective lepton number violating
$\znbb$ decay amplitude arises either by the $\Delta L=2$ (massive)
Majorana neutrino propagator (fig. 2.a), or through short-distance
four- and six-fermion lepton number violating vertices (fig. 2.b,c).
The latter appear in this model in two ways, either directly from the
trilinear parameters of the superpotential eq. \rf{WRPV}, or {\sl
indirectly} from the bilinear terms, due to the neutrino-neutralino
and chargino-charged lepton mixing.

From the superpotential one can read off the following $\Delta L=1$
trilinear terms which are relevant for $\znbb$ decay,

\ba{ltri} \nn
{\cal L}_{\rpm}^{(Trilinear)} & = & (\lambda_{d}^{ijk})^*
                      \Big\{ ({\bar \nu_i}P_R d_k){\tilde d}_{j,L}^{*}
                       -  ({\bar u_j}P_R d_k){\tilde e}_{i,L}^{*}
                        - ({\bar e_i}P_R d_k){\tilde u}_{j,L}^{*} \\
                      &  - &({\bar u_j}P_R e_i^C){\tilde d}_{k,R} \Big\} 
         + \lambda_{d}^{ijk}({\bar \nu_i^C}P_L d_j){\tilde d}_{k,R}^{*} 
                      + h.c.
\ea
As discussed above, the bilinear terms induce mixing between the SM
leptons and the MSSM charginos and neutralinos in the mass-eigenstate
basis, where one finds the following $\Delta L=1$ terms \cite{sensi},
\ba{lbi} \nn
{\cal L}_{\rpm}^{(Bilinear)} = & - & \frac{g}{\sqrt{2}}\kappa_n 
                      W_{\mu}^{-}({\bar e}\gamma^{\mu}P_L \chi_n^0) +\\ \nn
                      & \sqrt{2}g &
                      \Big\{\beta^u_k ({\bar \nu_k}P_R u^C){\tilde u}_{L}
                      +\beta^d_k ({\bar \nu_k}P_R d){\tilde d}_{R}^{*}
                      +\beta^e_{ki} ({\bar \nu_k}P_R e^C){\tilde e}_{Li} \\
                      & + & \beta^c({\bar u}P_R e^C){\tilde d}_{L}
                      \Big\}  + h.c.
\ea
The first term in eq. \rf{lbi} is generated from the Standard Model
$W$-boson-fermion-fermion and R-parity-conserving interactions of the
type $W-\chi^{0}-\chi^{\pm}$ present in the MSSM. The
R-parity-violating terms arise from the $\chi^{\pm}$-fermion-sfermion
and $\chi^{0}$-fermion-sfermion interactions.

The coefficients in eq. \rf{lbi} are defined as follows,

\be{alp}
\kappa_n = \sum_{k=1}^{3}\Delta^{-}_{1k}\Xi^{*}_{n+3,k} 
           + \sqrt{2} \Delta^{-}_{14}\Xi^{*}_{n+3,5}
           + \Delta^{-}_{15}\Xi^{*}_{n+3,6} 
\ee
\be{betd}
\beta^d_k = -\frac{1}{3}\tan\theta_W \Xi_{k,4}
\ee
\be{betu}
\beta^u_k =  -\frac{1}{6}(\tan\theta_W \Xi_{k,4} +3  \Xi_{k,5})
\ee
\be{bete}
\beta^e_{ki} =  -\frac{1}{\sqrt{2}} \Delta^{-}_{14} \Xi_{i,k}
               + \frac{1}{2}(\tan\theta_W \Xi_{k,4} + \Xi_{k,5})\delta_{i1}
\ee
\be{betc}
\beta^c = -\frac{1}{\sqrt{2}} \Delta^{-}_{14}.
\ee
Indices $i,k$ above run from $1$ to $3$, whereas the index $n=1,2,3,4$. 

If the \rp parameters are smaller than the other typical
supersymmetric parameters, one can find an approximate analytic
diagonalization for the rotation matrices $\Xi$ and $\Delta^{\pm}$
using the method introduced in ref.~\cite{774}.  This was done in
ref.~\cite{now96} and the results are presented in the appendix. In
the analysis presented later on we will use this convenient
approximative method and justify its use a posteriori. In leading
order in the expansion parameters $\xi$ and $\xi_L$ the coefficients
defined above are given by,

\be{appalp}
\kappa_n  \approx \sum_{m=1}^{4}N^{*}_{nm}\xi^{*}_{1m} 
           - \sqrt{2}(\xi_L)_{11}N^{*}_{n2} 
           - (\xi_L)_{12}N^{*}_{n3} 
\ee
\be{appbetd}
\beta^d_k  \approx \frac{1}{3} \tan\theta_W \sum_{j=1}^{3}
(V^{*}_\nu)_{jk}\xi^{*}_{j1}
\ee
\be{appbetu}
\beta^u_k  \approx \frac{1}{6} \sum_{j=1}^{3}(V_\nu^{*})_{jk}
             (\tan\theta_W\xi^{*}_{j1} + 3 \xi^{*}_{j2} )
\ee
\be{appbete}
\beta^e_{ki}  \approx \frac{1}{\sqrt{2}}(\xi_L)_{11}(V_\nu^{*})_{ki}
               -\frac{1}{2} \sum_{j=1}^{3}(V_\nu^{*})_{jk}
             (\tan\theta_W\xi^{*}_{j1} + \xi^{*}_{j2} ) \delta_{i1}
\ee
\be{appbetc}
\beta^c  \approx \frac{1}{\sqrt{2}}(\xi_L)_{11}
\ee
In the derivation we have always neglected terms which are of higher 
power in the neutrino mass and terms which are suppressed by 
factors of $m_{u,d,e}/M_{SUSY}$ due to left-right sfermion 
mixing and/or higgsino couplings. Here, $m_{u,d,e}$ are the masses 
of the up, down quarks and the electron mass and $M_{SUSY}$ is the 
typical mass scale of SUSY particles, assumed to be of order of 
${\cal O}(100$ GeV$)$. 

In addition to the lepton number violating Lagrangians eq. \rf{ltri}
and eq. \rf{lbi} for the calculation of ${\cal L}_{eff}$ for the
$\znbb$ decay the ordinary Standard Model (SM) charged-current
interaction as well as the MSSM gluino-quark-squark interaction

\be{laggluino}
{\cal L}_{\tilde g} = - \sqrt{2} g_3 
                        \frac{\lambda^{(a)}_{\alpha\beta}}{2}
       ({\bar q_{\alpha}}P_R {\tilde g}^{(a)} {\tilde q}_L^{\beta} - 
       {\bar q_{\alpha}}P_L {\tilde g}^{(a)} {\tilde q}_R^{\beta}) + h.c.
\ee
are needed. Here $\lambda^{(a)}_{\alpha\beta}$ are Gell-Mann matrices 
($a=1,...,8$), and the superscripts $\alpha$ and $\beta$ are color 
indices.

The leading order diagrams are shown in fig. 3. The first diagram,
fig. 3.a is nothing but the well-known mass mechanism, supplemented by
a contribution from neutralino exchange \footnote{Somewhat arbitrarily
we call the 4 heavy neutral states {\sl neutralinos}, while the light
states are referred to as {\sl neutrinos}. This is done, since in the
limit of vanishing \rp parameters the MSSM neutrino and neutralino
states are recovered.}. Since for finite values of the bilinear
parameters one neutrino acquires a tree-level mass this contribution
is always present in the bilinear \rp model. The diagram fig. 3.b
represents a short-range contribution to $\znbb$ decay induced by the
non-zero chargino-electron mixing. The remaining two diagrams are of
long-range type, and involve the standard {\sl Dirac-type} ($\Delta
L=0$) neutrino propagator. Note, that using the above approximations
these diagrams are always proportional to the product of a bilinear
times a trilinear parameter. However, these diagrams are also inherent
to the model, even in the basis where the trilinear parameters are set
to zero. Their contributions are, however, suppressed by a factor of
$m_f/m_{\tilde f}$ from left-right sfermion mixing, which should
naturally be very small for the first generation.  The diagrams in
fig. 3 do not include those proportional to $(\lambda_d^{ijk})^2$.
These have been previously analyzed in~\cite{hir9598}.

Integrating out the heavy fields and carrying out a Fierz 
transformation, one obtains finally the effective Lagrangian 

\ba{leff} \nn
{\cal L}_{eff}(x) &=& 
 \sqrt{2}G_F (V_{\nu})_{1k}({\bar e}\gamma_{\mu}P_L \nu_k) J^{\mu}  \\
& + &\frac{G_F^2}{2 m_p}
                    (\eta_{\tilde g}+\eta_{\chi}) J^{\mu}J_{\mu} 
                    ({\bar e}P_R e^c) - \\ \nn
 &-& \sqrt{2}G_F \Big( \eta_{\lambda}^{(k)} ({\bar \nu_k}P_R e^c) J 
  + \eta_{\lambda,q}^{(k)} ({\bar \nu_k}\sigma_{\mu\nu}P_R e^c)
                            J^{\mu\nu}\Big) 
 \ea
The first term in the above equation is the usual SM charged-current
interaction. Color-singlet hadronic currents in eq. \rf{leff} are
given by

\be{defsc}
J = {\bar u}^{\alpha} P_R d_{\alpha},
\ee
\be{defvac}
J^{\mu} = {\bar u}^{\alpha} \gamma_{\mu}P_L d_{\alpha},
\ee
\be{deftc}
J^{\mu\nu} = {\bar u}^{\alpha} \sigma^{\mu\nu}P_R d_{\alpha}.
\ee
In eq. \rf{leff} we have introduced effective lepton number 
violating parameters $\eta$ defined by,

\be{etagl}
\eta_{\tilde g}=\Big(\frac{4\pi\alpha_S}{9}\Big)
                \Big(\frac{4\pi\alpha_2}{G_F^2 m_{\tilde d_L}^4}\Big)
                \Big(\frac{m_p}{m_{\tilde g}}\Big) (\beta^c)^2
\ee
\be{etac}
\eta_{\chi}=\sum_{i=1}^4 \frac{m_p}{m_{\chi^0_{i}}}\kappa_i^2 
           =: \frac{m_p}{\langle m_{\chi}\rangle}
\ee
\be{etal}
\eta_{\lambda}^{(k)} = \frac{g\lambda_{d}^{i11}}{2G_F}\Big\{
                 2\frac{\beta^{e}_{ki}}{m_{\tilde e_{L,i}}^2} -
                  \frac{\beta^{u}_{k}}{m_{\tilde u_L}^2}\delta_{i1} -
                  \frac{\beta^{d}_{k}}{m_{\tilde d_R}^2}\delta_{i1} +
                  \frac{\beta^{c}}{m_{\tilde d_L}^2}\delta_{ik} \Big\}
\ee
\be{etaq}
\eta_{\lambda,q}^{(k)} = \frac{g\lambda_{d}^{i11}}{8G_F}\Big\{
                  \frac{\beta^{u}_{k}}{m_{\tilde u_L}^2}\delta_{i1} -
                  \frac{\beta^{d}_{k}}{m_{\tilde d_R}^2}\delta_{i1} +
                  \frac{\beta^{c}}{m_{\tilde d_L}^2}\delta_{ik} \Big\}
\ee
Here, $G_F$ is the Fermi constant and $m_p$ is the proton mass.

At this point, we would like to discuss some differences between our
calculation and an earlier work \cite{FKS98}. In ref. \cite{FKS98} it
was found that the short-ranged part due to gluino exchange is 
proportional to pseudo-scalar and tensor hadronic currents. (This 
would imply an important enhancement of the gluino
diagram due to the pion exchange mechanism first discussed for 
the case of trilinear R-parity breaking in \cite{fae97}.) However, we 
find that the
gluino-mediated diagram is proportional to the same hadronic structure
as appears in the heavy Majorana neutrino exchange, characteristic of
left-right symmetric models \cite{doi85,lrm}.  Therefore, we conclude 
that the gluino contribution is negligibly small. Also in the long-range 
part of the effective Lagrangian, eq. \rf{leff}, the last term in 
eq. \rf{etal} is missing in the corresponding equation in ref. \cite{FKS98}, 
and the long-range tensor contribution proportional to 
$\eta_{\lambda,q}^{(k)}$ has been neglected compared to the long-range
scalar-pseudoscalar part in \cite{FKS98}.

The effective Lagrangian given above is defined in terms of quark
currents. For $\znbb$ decay we have to reformulate ${\cal L}_{eff}$ in
terms of nucleon currents and apply the proper nuclear wave functions
to describe nucleon states in the nucleus. This is done in the
following section.

\section{Half-life of $\znbb$ decay and definitions of nuclear 
matrix elements}

Having specified the effective Lagrangian eq. \rf{leff} the 
derivation of the half-life of double beta decay is rather 
straightforward. Following the standard procedure described in 
\cite{hir9598,hir98}, after lengthy manipulations one finds,

\be{deft12}
\Big(T_{1/2}^{\znbb}\Big)^{-1} = 
G_{01}({\cal M}_{\nu})^2 |\eta^{\Delta L=2}|^2.
\ee
Here, $G_{01}$ is the leptonic phase space integral, numerical values
can be found in \cite{doi85}, ${\cal M}_{\nu}$ is the nuclear matrix
element governing the well-known mass mechanism of double beta decay
and

\be{defeta}
\eta^{\Delta L=2} = \frac{\emass}{m_e} + 
                    (\frac{m_P}{\langle m_{\chi} \rangle} + 
                     \eta_{\tilde g}) {\cal R}_{N}
                  + \eta_{\lambda}{\cal R}_{\lambda}
                  + \eta_{\lambda,q}{\cal R}_{\lambda,q},
\ee    
where ${\cal R}_j = {\cal M}_j/{\cal M}_{\nu}$ with $j=\lambda,
\lambda q,N$.

Individual nuclear matrix elements are defined by,

\be{mnu}
{\cal M}_{\nu}=(M_{GT,m}-(\frac{g_V}{g_A})^2 M_{F,m}),
\ee
\be{mnuh}
{\cal M}_{N}=(M_{GT,N}-(\frac{g_V}{g_A})^2 M_{F,N}),
\ee
\be{mlam}
{\cal M}_{\lambda}=\alpha_{1}(\frac{1}{3}M_{GT'}+M_{T'}),
\ee
\be{mlamq}
{\cal M}_{\lambda,q}=(-\frac{2}{3}\alpha_{2}M_{GT'}+\alpha_{3}M_{F'}+
                     \alpha_{2}M_{T'}).
\ee
The coefficients $\alpha_i$ are given by \cite{LIBB}

\be{defa1}
\alpha_1 = \frac{F_P^{(3)}}{(Rm_e)g_A},
\ee
\be{defa2}
\alpha_2 = \frac{4T_1^{(3)}g_V(1-2m_p(g_W/g_V))}{g_A^2 (Rm_e)},
\ee
\be{defa3}
\alpha_3 = \frac{4(2{\hat T}_2^{(3)}-T_1^{(3)})g_V}{g_A^2 (Rm_e)},
\ee
with $g_A \simeq 1.26$, $g_V \simeq 1$ and $(g_W/g_V) \simeq
-3.7/(2m_P)$ from CVC, while numerical values for $T_i^{(3)}$ can be
found in ref. \cite{adler}. $m_e$ and $R$ are the electron mass and
the nuclear radius. The matrix elements in the closure approximation
are (summation over nucleon indices are suppressed) given by
\cite{mut89,LIBB}

\be{matrgt}
M_{GT,k}=
\langle 0_f^+| h_{k}(\vec{\sigma_a}
\vec{\sigma_b})\tau^+_a \tau^+_b| 0_i^+ \rangle 
\ee
\be{matrf}
M_{F,k}=
\langle 0_f^+| h_{k}\tau^+_a \tau^+_b| 0_i^+ \rangle 
\ee
\be{matrtp}
M_{T^{'}}=\langle 0_f^+| h_{T^{'}}\{
(\vec{\sigma_a}\hat{\vec{r}}_{ab})(\vec{\sigma_b}\hat{\vec{r}}_{ab})
-\frac{1}{3}(\vec{\sigma}_a\vec{\sigma}_b)\} 
\tau^+_a \tau^+_b| 0_i^+ \rangle.
\ee
with $k=m,R,N$ and $h_k$ and $h_{T^{'}}$ are neutrino potentials 
defined as
\be{nepotmass}
h_m =\frac{2}{\pi}R\int_0^\infty dq q^2
\frac{j_0(qr_{ab})f^2(q^2)}{\omega(\omega+\overline{E})}, 
\ee
\be{nepot01}
h_R =\frac{2}{\pi}\frac{R^2}{m_P}\int_0^\infty dq q^4
\frac{j_0(qr_{ab})f^2(q^2)}{\omega(\omega+\overline{E})}, 
\ee
\be{nepot02}
h_{T^{'}} = \frac{2}{\pi}\frac{R^2}{m_P}\int_0^\infty dq q^2
\frac{f^2(q^2)}{\omega(\omega+\overline{E})}\{q^2 j_0(qr_{ab}) 
-3 \frac{q}{r_{ab}} j_1(qr_{ab})\}. 
\ee
\ba{F_N}
h_N (x_A) &=& \frac{4 \pi m_{A}^{8}}{m_p m_e}  r_{ab} \int
\frac{d^3{\bf q}}{(2\pi)^3} \frac{1}{(m_A^2 + {\bf q}^2)^4}
        e^{i {\bf q} {\bf r}_{ab}}
\ea
with $x_A = m_A r_{ab}$, $\omega = \sqrt{{\bf q}^2 + m_{\nu}^2}$, 
$\overline{E}$ the mean excitation of the intermediate nuclear 
states and $r_{ab}$ the nucleon-nucleon distance. $f^2(q^2)$ is 
the nucleon form factor, usually taken to have dipole form, $j_0$ 
and $j_1$ are spherical Bessel functions, and $m_A$ is  
typically $m_A = 0.85$ GeV \cite{mut89}.

\section{$\znbb$ Decay Constraints on \rp Parameters}

Up to now our calculation has been independent of the use of any
nuclear structure model. However, for a numerical analysis we have to
assume some definite numbers for the relevant nuclear matrix
elements. For definiteness we will use the pn-QRPA values of the
nuclear matrix elements \cite{hir9598,lrm,LIBB,mut89}. Note that the
uncertainty of limits on the fundamental \rp parameters depend only on
the square roots of the uncertainties of the nuclear structure matrix
elements.

In the following analysis we will use the currently most stringent 
limit on double beta decay 

\be{texp}
T_{1/2}(^{76}Ge) \ge 1.1 \times 10^{25} ys
\ee
found by the Heidelberg-Moscow experiment \cite{hdmo}. 

The experimental constraint eq. \rf{texp} together with the theoretical 
decay rate eq. \rf{deft12} defines an excluded area in a complex 
multi-dimensional parameter space consisting of the \rp parameters 
as well as a number of MSSM parameters. 

We have analyzed the contributions from the different Feynman
diagrams, see fig. 3, and found that
%
for typical choices of MSSM parameters the neutrino mass
mechanism is always the dominant contribution. To understand this
finding, let us first explicitly calculate the effective neutrino mass
in the bilinear \rp model, which, using the usual GUT assumption
\footnote{Although we will use this assumption in the following, 
our results do not sensitively depend on this choice.}  $M_1 =
\frac{5}{3}\tan\theta_W^2 M_2$, is given by,

\be{defmeff}
\langle m_{\nu} \rangle = \sum_j' U_{ej}^2 m_j 
= \frac{2}{3} \frac{g^2 M_2}{det({\cal M}_{\chi^0})}\Lambda_1^2
\ee
where $\Lambda_1 = \epsilon_1 v_1 - \omega_1 \mu$ (see appendix) and
the prime indicates summation only over light fermion states (not
neutralinos), and $\omega_1 := \langle {\tilde \nu_1}\rangle$.

Let us first compare the short-range contributions with the long-range
(light) neutrino exchange. All of them are of the same order in RPV
parameters, i.e. proportional to $\Lambda_1^2$.  By naively comparing
the corresponding diagrams one might think that they are of equal
importance. Closer inspection, however, reveals that this is not the
case, and that the short-range contributions are suppressed compared
to the long-range ones.  This suppression is due to nuclear physics
effects. For heavy particle exchange the two quarks interacting have
to come very close together, where the strong repulsive part of the
nucleon-nucleon interaction is important. For point-like nucleons this
contribution would be zero, so that the corresponding nuclear matrix
elements become non-zero only due to the finite nucleon size.  A naive
estimate of the size of the short-range contributions relative to the
neutrino mass contribution can be given by comparing the typical
momentum scale of the nucleons, order $p_F$, with the typical mass
scale of SUSY particles $m_{SUSY}$ and thus is expected to be
$p_F/m_{SUSY} \sim 10^{-3}$.  Numerically we have found that indeed
the constraints on RPV parameters derived from the gluino diagram
alone are weaker than the constraints derived from considering light
neutrino exchange by about the above factor.

In contrast, for the long-range diagrams, fig 3.c and 3.d, an
enhancement compared to the neutrino mass diagram would be naively
expected. The explicit calculation, however, shows that these diagrams
are proportional to $\lambda_d^{i11}\Lambda_1$ (in a basis with
non-zero trilinear terms) or proportional to $m_f/m_{\tilde
f}\Lambda_1 \simeq {\cal O}(10^{(-6)})\Lambda_1$ for first generation
fermions (in a basis where $\lambda_d^{i11}$ is assumed to be
zero). Thus despite a huge enhancement factor of these diagrams, from
the propagation of the neutrino {\sl a la Dirac}, no useful limits on
the fundamental parameters $\epsilon_1$ and $\omega_1$ can be derived
from these contributions.

Fig. 4 then shows the limits on the parameters $\epsilon_1$ (fig. 4.a) 
and $\omega_1$ (fig. 4.b), assuming $\tan\beta=1$ as a function 
of the MSSM parameters ($M_2,\mu$). Limits typically of the order 
of a few hundred keV are found, choosing $M_2$ and $\mu$ of the order 
of up to a few hundred GeV. Note, however, that the limit on 
$\epsilon_1$ depends rather strongly on the choice of $\tan\beta$, 
and detoriates proportional to $\tan\beta$ for large values 
of this parameter. 

%
%
The robustness of the double beta decay bound is then demonstrated 
in fig. 5, where calculated half-lives for the $\znbb$ decay 
of $^{76}Ge$ as a function of $\langle {\tilde \nu_1}\rangle$ 
are shown. For this figure we have varied the MSSM parameters 
freely and randomly in a range motivated by naturalness arguments, i.e. 
$M_2$ and $\mu$ between $100$ GeV and $1$ TeV and $\tan\beta = 1-50$. 
A clear absolute upper limit of the order of $1.5$ MeV on
$\langle {\tilde \nu_1}\rangle$ is found from this analysis.

The limits shown in fig. 4, 5 assume that only one bilinear \rp parameter
is non-zero, which is unnatural from the point of view of the
minimization of the scalar potential~\cite{epsrad}. A combined limit
on ($\epsilon_1, \omega_1$) is therefore shown in fig. 6., for $\mu =
100$ GeV, various values of $M_2$ and $\tan\beta=1$ (fig. 6.a) and
$\tan\beta=10$ (fig. 6.b). Obviously, the limits on $\epsilon_1$
depend rather strongly on $\tan\beta$, whereas the limits on
$\omega_1$ do not. Note the double logarithmic scale.

%
%
%

In summary we find that $\znbb$ decay provides very stringent
limits on RPV parameters of the first generation. For any reasonable
choice of MSSM parameters in the bilinear \rp model the mass mechanism
is the dominant contribution to $\znbb$ decay.


\section{Discussion and Conclusion}

We have calculated the contributions of (bilinear) R-parity breaking
supersymmetry to neutrinoless double beta decay.  This model mimics
closely the case of spontaneous breaking of R-parity. We find that
$\znbb$ decay constrains only a subset of the possible bilinear
parameters, namely $\epsilon_1$ and the sneutrino VEV of the first
generation $\omega_1$. This is a general property of the theory and
does not mean any fine-tuning of parameters.  For the first generation
\rp parameters, on the other hand, $\znbb$ decay provides very
stringent limits, typically of the order of a few hundred keV up to a
few MeV. With these limits at hand it seems rather hopeless to search
for RPV in the first generation at colliders. From this point of view
a possible interpretation of HERA events in terms of \rp interactions
would be rather unlikely. However, it is important to stress that,
even though we find rather stringent constraints on the magnitude of
first generation \rp parameters, these do not limit in any way the
attainable magnitudes of R-parity breaking signatures expected at
colliders, since the latter involve mainly the third generation,
i.e. they involve $\tau$'s or \nt's. This fact follows very naturally,
for example, in scenarios with radiative symmetry breaking, due to the
larger Yukawa couplings that characterize the third generation.

In our present study we find that the main origin $\znbb$ decay in
this model is the mass mechanism and conclude that other contributions
are practically irrelevant. 

\centerline{\bf Acknowledgement}

We are grateful to S.G. Kovalenko for useful comments. This work was
supported by the Spanish DGICYT under grant PB95-1077 and by the
European Union's TMR program under grants ERBFMRXCT960090 and
ERBFMBICT983000.

\section{Appendix A: Mass matrices}

\subsection{Neutral fermion mass matrix}

The presence of the lepton number violating bilinear terms in the
superpotential, see eq. \rf{superpot}, leads to mixing of the
neutralino and neutrino states~\cite{oldRPV}. We treat the effects of
R-parity violation perturbatively, using the method in~\cite{774} in
order to diagonalize the various mass matrices and hence determine the
corresponding couplings. In doing this we follow the conventions given
in ref. \cite{now96}.  In two component notation let $\Psi'_{0}$
denote the column vector of neutrinos and neutralinos in the basis
\be{nwbas}
{\Psi'_{0}}^T = (\psi_{L_1}^1, \psi_{L_2}^1, \psi_{L_3}^1, 
                 -i\lambda', -i \lambda_3, 
                 \psi_{H_1}^1,\psi_{H_2}^2),
\ee
where $\psi_{L_i}^1$ are the neutrino fields, $-i\lambda'$ and 
$-i \lambda_3$ are the unmixed photino and gaugino states, 
respectively, and $\psi_{H_{1,2}}^{1,2}$ are the neutral components of 
the Higgs doublets. The Lagrangian describing the neutrino/neutralino 
masses is then given by,

\be{nmlag}
{\cal L}^0_{mass} = -\frac{1}{2}{\Psi'_{0}}^T {\cal M}_0 {\Psi'_{0}} 
                    + {\rm h.c.}
\ee
The mass matrix can be written in a block form,

\be{nmm}
{\cal M}_0 =  \left(
                    \begin{array}{cc}
                    0 & m \\
                    m^T & {\cal M}_{\chi^0} \\
                    \end{array}
              \right).
\ee
Here, the submatrix $m$ contains entries from the bilinear \rp 
parameters,

\be{bnmm}
m =   \left(
            \begin{array}{cccc}
     -\frac{1}{2}g'\omega_e & \frac{1}{2}g\omega_e & 0 & -\epsilon_e \\
-\frac{1}{2}g'\omega_{\mu} & \frac{1}{2}g\omega_{\mu} & 0 & -\epsilon_{\mu} \\
        -\frac{1}{2}g'\omega_{\tau} & \frac{1}{2}g\omega_{\tau} & 
          0 & -\epsilon_{\tau} \\
                    \end{array}
              \right),
\ee
$\omega_i := \langle {\tilde \nu}_{i} \rangle$. ${\cal M}_{\chi^0}$ 
is the MSSM neutralino mass matrix, given by,

\be{MSSMnm}  
{\cal M}_{\chi^0} =  \left(
                        \begin{array}{cccc}
 M_1 & 0   & -\frac{1}{2}g' v_1 &  \frac{1}{2} g' v_2  \\
 0   & M_2 &  \frac{1}{2}g  v_1 & -\frac{1}{2} g  v_2  \\
  -\frac{1}{2} g' v_1 &  \frac{1}{2} g v_1 & 0 & -\mu  \\
   \frac{1}{2} g' v_2 & -\frac{1}{2} g v_2 & -\mu & 0 \\
 \end{array}
                     \right).
\ee
One defines the mass eigenstates by means of the rotation

\be{m0eigen}
\Psi_{0,i} = \Xi_{ij}\Psi'_{0,j}.
\ee
The mass matrix then is diagonalize by 
\be{d0nm}
\Xi^{*} {\cal M}_0 \Xi^{\dagger} = {\hat{\cal M}_0},
\ee
where ${\hat{\cal M}_0}$ is a diagonal matrix, with the eigenvalues 
as elements. Note that these conventions are chosen such that in the 
limit of vanishing \rp parameters the usual MSSM convention \cite{HK85}, 
$\Xi \rightarrow {\cal N}$, is recovered.

\subsection{Charged fermion mass matrix}

For the charged fermion sector the mass part of the Lagrangian can be
written as,

\be{lcmass}
{\cal L}^{\pm}_{mass} = - \Psi_{(-)}'^T {\cal M}_{\pm} 
                          \Psi_{(+)}' + h.c.
\ee
where in two component Weyl spinor basis

\be{psipminus}
 \Psi_{(-)}'^T = (\psi^{2}_{L_1},\psi^{2}_{L_2},\psi^{2}_{L_3},
                 -i\lambda_{-},\psi_{H_1}^{2}),
\ee
\be{psipplus}
 \Psi_{(+)}'^T = (\psi_{R_1}^c,\psi_{R_2}^c,\psi_{R_3}^c,-i\lambda_{+},
                 \psi_{H_2}^{1}).
\ee
The ($5\times 5$) charged fermion mass can be written as, 

\be{defcm}
{\cal M}_{\pm} =  \left(
                    \begin{array}{cc}
                    M_l & E \\
                    E' & {\cal M}_{\chi^{\pm}} \\
                    \end{array}
              \right),
\ee
where ${\cal M}_{\chi^{\pm}}$ is the usual MSSM chargino mass matrix 

\be{defmssmc}
{\cal M}_{\chi^{\pm}} =  \left(
                    \begin{array}{cc}
                    M & \frac{1}{\sqrt{2}}g v_1 \\
                     \frac{1}{\sqrt{2}}g v_2 & \mu \\
                    \end{array}
              \right),
\ee
and the sub-matrices $E$ and $E'$ give rise to chargino-charged lepton 
mixing. They are defined as

\be{defE}
E =   \left(
                    \begin{array}{cc}
                    \frac{1}{\sqrt{2}}g \omega_e & \epsilon_e \\
                    \frac{1}{\sqrt{2}}g \omega_{\mu} & \epsilon_{\mu} \\
                    \frac{1}{\sqrt{2}}g \omega_{\tau} & \epsilon_{\tau} \\
                    \end{array}
              \right),
\ee
\be{defEp}
E' =   \left(
             \begin{array}{ccc}
             0 & 0 & 0 \\
             \frac{1}{\sqrt{2}} h_{1i} \omega_i & 
             \frac{1}{\sqrt{2}} h_{2i} \omega_i & 
             \frac{1}{\sqrt{2}} h_{3i} \omega_i \\
                    \end{array}
              \right).
\ee
$M_l$ finally is the charged lepton mass matrix due to the ordinary 
Yukawa interactions. The rotation from the weak to the mass 
eigenstate basis is defined as,

\be{rotc}
\Psi_{(\pm)i} = \Delta^{\pm}_{ij}\Psi'_{(\pm)j},
\ee
where $\Delta^{\pm}$ diagonalize the mass matrix via 

\be{diagc}
(\Delta^{-})^{*} {\cal M}_{\pm} (\Delta^{+})^{\dagger} = 
Diag\{m_{i}^{l},m_{\chi^{\pm}_k}\}.
\ee
$m_{i}^{l}$ are the known masses of the charged leptons, and 
$m_{\chi^{\pm}_k}$ the ``chargino'' masses, $k=1,...,4$.

Note, that without making assumptions about the structure of $M_l$, 
it is impossible to diagonalize ${\cal M}_{\pm}$. For numerical 
purposes one therefore has to assume that $M_l$ (i.e. the Yukawa 
couplings $h_{ij}$) can be brought into a diagonal form. 
The constraint that the smallest 3 eigenvalues of ${\cal M}_{\pm}$ 
should coincide with the known masses of the charged leptons, 
can than be reformulated as a set of equations for the Yukawa 
couplings, which for finite \rp parameters differ from the MSSM 
equations.

\section{Appendix B: Approximate diagonalization of mass matrices}

\subsection{Neutral fermion mass matrix}

If the \rp parameters are small in the sense that for 
\be{exp0}
\xi = m \cdot {\cal M}_{\chi^0}^{-1}
\ee
all $\xi_{ij} \ll 1$, one can find an approximate solution for 
the neutrino/neutralino mass matrix $\Xi$. Explicitly, the 
$\xi_{ij}$ are given by 

\be{xi1}
\xi_{i1} = \frac{g' M_2 \mu}{2 det({\cal M}_{\chi^0})}\Lambda_i,
\ee
\be{xi2}
\xi_{i2} = -\frac{g M_1 \mu}{2 det({\cal M}_{\chi^0})}\Lambda_i,
\ee
\be{xi3}
\xi_{i3} = \frac{\epsilon_i}{\mu} + 
          \frac{(g^2 M_1 + {g'}^2 M_2) v_2}
               {4 det({\cal M}_{\chi^0})}\Lambda_i,
\ee
\be{xi4}
\xi_{i4} = - \frac{(g^2 M_1 + {g'}^2 M_2) v_1}
               {4 det({\cal M}_{\chi^0})}\Lambda_i.
\ee
It is interesting to note, that only $\xi_{i3}$ is {\it not} 
proportional to $\Lambda_i$, where $\Lambda_i$ is 

\be{deflam}
\Lambda_i = \mu \omega_i - v_1 \epsilon_i.
\ee
In leading order in $\xi$ the mixing matrix $\Xi$ is given by,
\ba{xiapprox}
\Xi^* & = & \left(\begin{array}{cc}
V_\nu^T & 0 \\
0 & N^* \end{array}\right)
\left(\begin{array}{cc}
1 -{1 \over 2}\xi \xi^{\dagger} & -\xi \\
\xi^{\dagger} &  1 -{1 \over 2}\xi^\dagger \xi
\end{array}\right) \\ \nn
& = &  \left(\begin{array}{cc}
V_\nu^T(1 -{1 \over 2}\xi \xi^{\dagger} ) & -V_\nu^T\xi \\
N^* \xi^{\dagger} & N^* ( 1 -{1 \over 2}\xi^\dagger \xi) 
\end{array}\right)
\ea
The second matrix in eq. \rf{xiapprox} above block-diagonalize 
${\cal M}_0$ approximately to the form 
diag($m_{eff},{\cal M}_{\chi^0}$), where

\be{meff}
m_{eff} = - m \cdot {\cal M}_{\chi^0}^{-1} m^T = 
\frac{M_1 g^2 + M_2 {g'}^2}{4 det({\cal M}_{\chi^0})} 
\left(\begin{array}{ccc}
\Lambda_e^2 & \Lambda_e \Lambda_\mu
& \Lambda_e \Lambda_\tau \\
\Lambda_e \Lambda_\mu & \Lambda_\mu^2
& \Lambda_\mu \Lambda_\tau \\
\Lambda_e \Lambda_\tau & \Lambda_\mu \Lambda_\tau & \Lambda_\tau^2
\end{array}\right).
\ee
Here, $det({\cal M}_{\chi^0})$ is the determinant of ${\cal M}_{\chi^0}$ 

\be{detmc}
det({\cal M}_{\chi^0}) = - \mu^2 M_1 M_2 + \frac{1}{2} \mu v_1 v_2
                           (M_1 g^2 + M_2 {g'}^2).
\ee
The submatrices $N$ and $V_{\nu}$ in eq. \rf{xiapprox} diagonalize 
${\cal M}_{\chi^0}$ and $m_{eff}$ in the following way:

\be{defdmssm}
N^{*}{\cal M}_{\chi^0} N^{\dagger} = diag(m_{\chi^0_i}),
\ee
\be{defdn}
V_{\nu}^T m_{eff} V_{\nu} = diag(0,0,m_{\nu}),
\ee
where $m_{\chi^0_i}$ are the ``neutralino'' mass eigenstates, and 
the only non-zero neutrino mass is given by 
\be{nonzero}
m_{\nu} = Tr(m_{eff}) = 
\frac{M_1 g^2 + M_2 {g'}^2}{4 det({\cal M}_{\chi^0})} 
|{\vec \Lambda}|^2.
\ee
Note that this convention is chosen such that eq. \rf{defdmssm}
coincides with the standard MSSM convention by Haber and Kane
\cite{HK85}.

For $V_{\nu}$ one can find an analytic form, 

\be{vana}
V_{\nu} = 
\left(\begin{array}{ccc}
  1 &                0 &               0 \\
  0 &  \cos\theta_{23} & -\sin\theta_{23} \\
  0 &  \sin\theta_{23} & \cos\theta_{23} 
\end{array}\right) \cdot
\left(\begin{array}{ccc}
  \cos\theta_{13} & 0 & -\sin\theta_{13} \\
                0 & 1 &               0 \\
  \sin\theta_{13} & 0 & \cos\theta_{13} 
\end{array}\right) ,
\ee
i.e.,
\be{vtana}
V_{\nu} = 
\left(\begin{array}{ccc}
  \cos\theta_{13} & 0 & -\sin\theta_{13} \\
 -\sin\theta_{13}\sin\theta_{23} &  \cos\theta_{23} & 
-\cos\theta_{13}\sin\theta_{23} \\
  \sin\theta_{13}\cos\theta_{23} & \sin\theta_{23} & 
\cos\theta_{13}\cos\theta_{23} 
\end{array}\right) ,
\ee
where the mixing angles can be expressed in terms of the 
``alignment vector'' ${\vec \Lambda}$ as follows:

\be{th13}
\tan\theta_{13} = - \frac{\Lambda_e}
                   {(\Lambda_{\mu}^2+\Lambda_{\tau}^2)^{\frac{1}{2}}},
\ee
\be{th23}
\tan\theta_{23} = \frac{\Lambda_{\mu}}{\Lambda_{\tau}}.
\ee

\subsection{Charged fermion mass matrix}

Similarly to the case of the neutralino mass matrix discussed above, 
for small values of the \rp breaking parameters it is possible 
to find an approximate diagonalization procedure for the charged 
mass matrix. 

Using the fact that the matrix $E'$ is proportional to the Yukawa 
couplings of the charged leptons, one can assume $E' \approx 0$. 

Define then,

\be{xil}
\xi_L^{*} = E ({\cal M}_{\chi^{\pm}})^{-1},
\ee
\be{xir}
\xi_R^{*} = M_l^{\dagger} E {\cal M}_{\chi^{\pm}}^{-1}
({\cal M}_{\chi^{\pm}}^{-1})^T = 
M_l^{\dagger} \xi_L^{*}({\cal M}_{\chi^{\pm}}^{-1})^T.
\ee
Note that $\xi_R \sim \xi_L (m_l/M)$. Explicitly,

\be{defxil1}
(\xi_L^{*})_{i1} =\frac{g \Lambda_i}
                  {{\sqrt{2}}(M\mu-\frac{1}{2}g^2v_1v_2)} 
\ee
\be{defxil2}
(\xi_L^{*})_{i2} = \frac{\epsilon_i}{\mu} 
                 -\frac{g^2 v_2\Lambda_i}
                  {2 \mu(M\mu-\frac{1}{2}g^2v_1v_2)} 
\ee
The rotation matrices $\Delta^{\pm}$ are then given in leading order 
in $\xi_L$ and $\xi_R$ as 

\be{apprdm}
(\Delta^{-})^{*}  =  \left(\begin{array}{cc}
V_L & 0 \\
0 & U^* \end{array}\right)
\left(\begin{array}{cc}
1 -{1 \over 2}\xi_L^{*} \xi_L^{T} & -\xi_L^{*} \\
\xi_L^{T} &  1 -{1 \over 2}\xi_L^T \xi_L^{*}
\end{array}\right) ,
\ee
\be{apprdp}
(\Delta^{+})^{\dagger}  =
\left(\begin{array}{cc}
1 -{1 \over 2}\xi_R^{*} \xi_R^{T} & \xi_R^{*} \\
-\xi_R^{T} &  1 -{1 \over 2}\xi_R^T \xi_R^{*}
\end{array}\right)
\left(\begin{array}{cc}
V_R^{\dagger} & 0 \\
0 & V^{\dagger} \end{array}\right).
\ee
The conventions are chosen such, that in the limit of vanishing 
\rp eqs \rf{apprdm},\rf{apprdp} simplify to the MSSM notations:

\be{diagcm}
U^{*} {\cal M}_{\chi^{\pm}} V^{\dagger} = {\hat {\cal M}}_{\chi^{\pm}},
\ee
\be{diagl}
V_L M_l V_R^{\dagger}  = {\hat M_l},
\ee
where again the hat indicates a diagonal matrix.

%
%
\hskip25mm
\epsfysize=60mm
\epsfxsize=80mm
\epsfbox{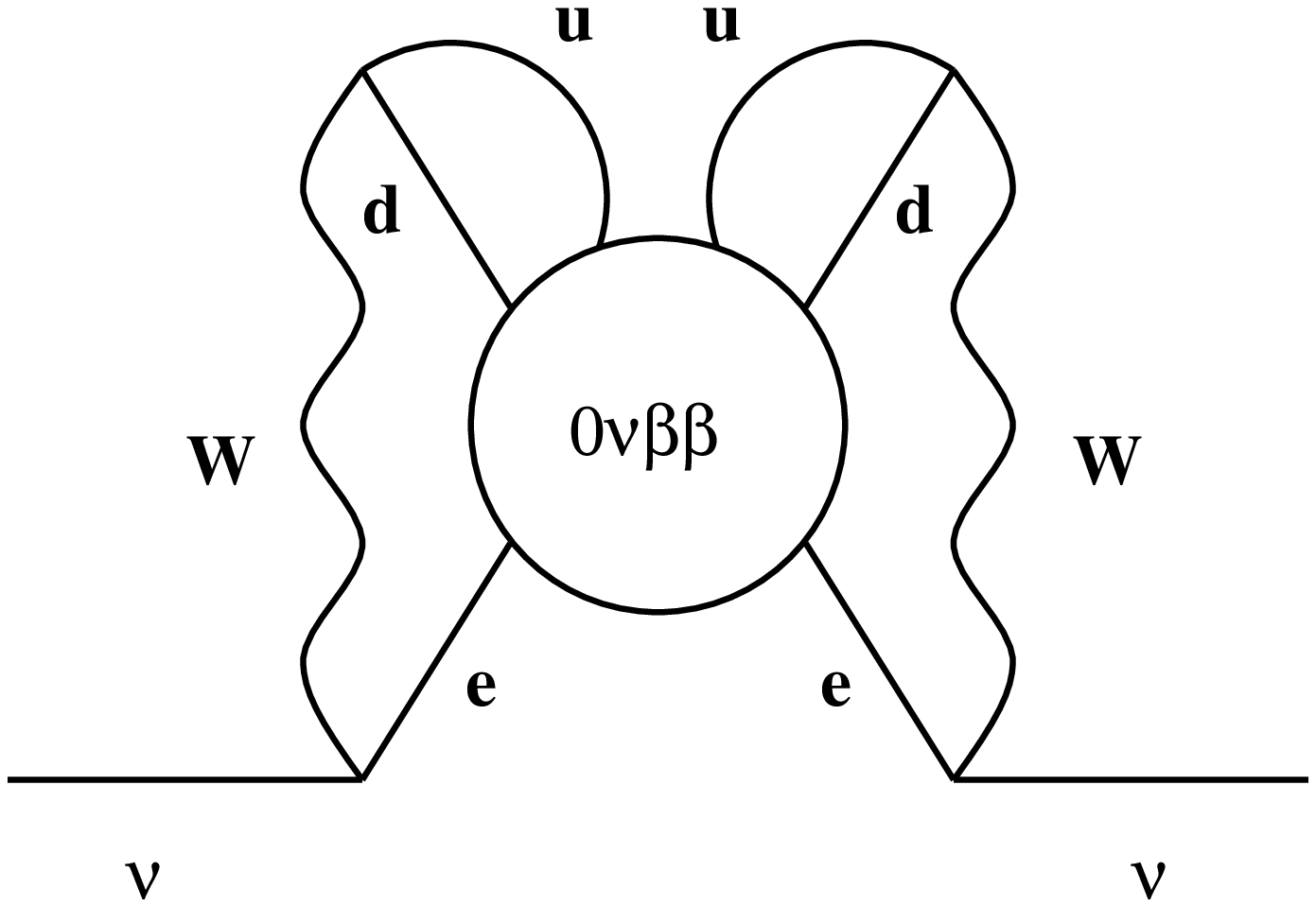}

\bigskip
\noindent
{\bf Figure 1: }{\it Diagram illustrating the connection between the 
Majorana mass of the neutrino and the amplitude of double beta decay.}

\bigskip

%
%
\vskip20mm
\epsfysize=40mm
\epsfxsize=140mm
\epsfbox{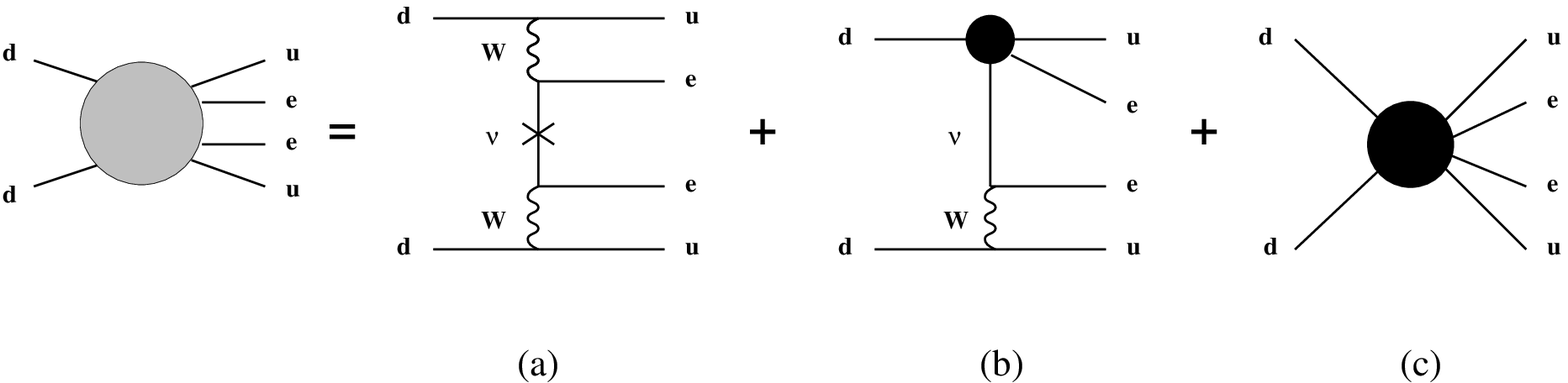}

\bigskip
\noindent
{\bf Figure 2: }{\it Decomposition of the double beta decay amplitude 
into long-range and short-range contributions.}
\bigskip

%
%
\vskip05mm
\epsfysize=50mm
\epsfxsize=50mm
\epsfbox{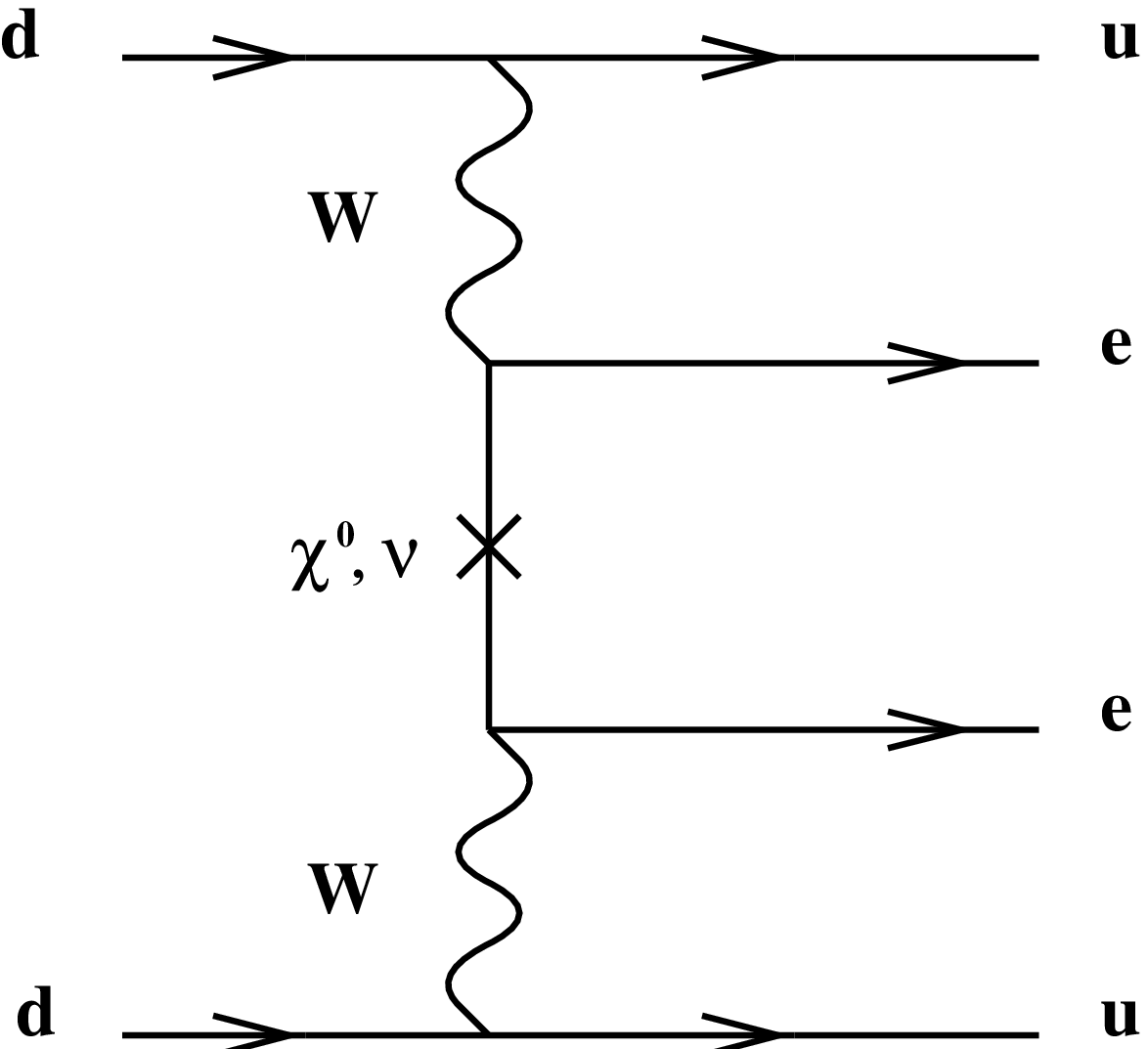}

\vskip-50mm
\hskip60mm
\epsfysize=50mm
\epsfxsize=50mm
\epsfbox{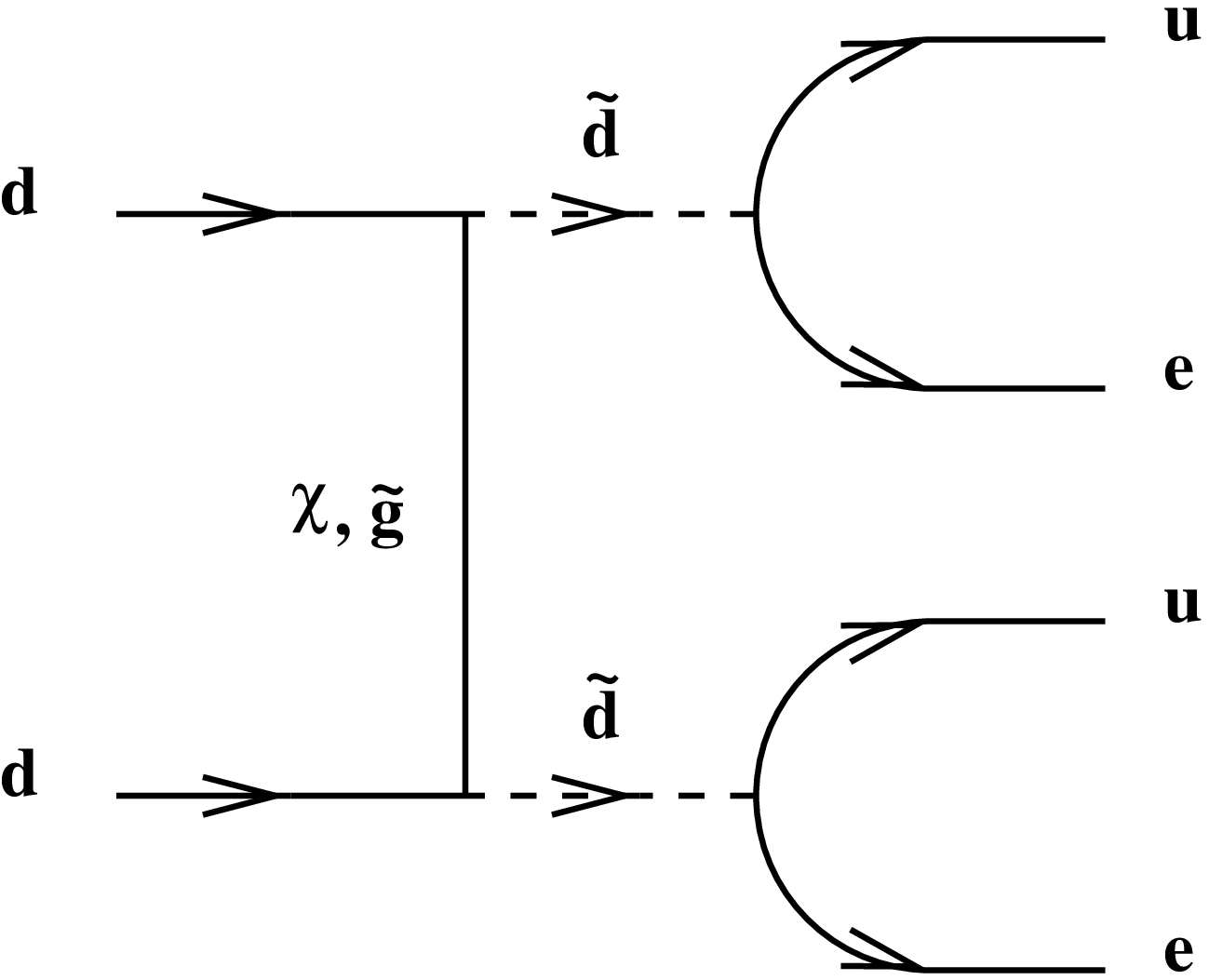}

\vskip05mm
\epsfysize=50mm
\epsfxsize=50mm
\epsfbox{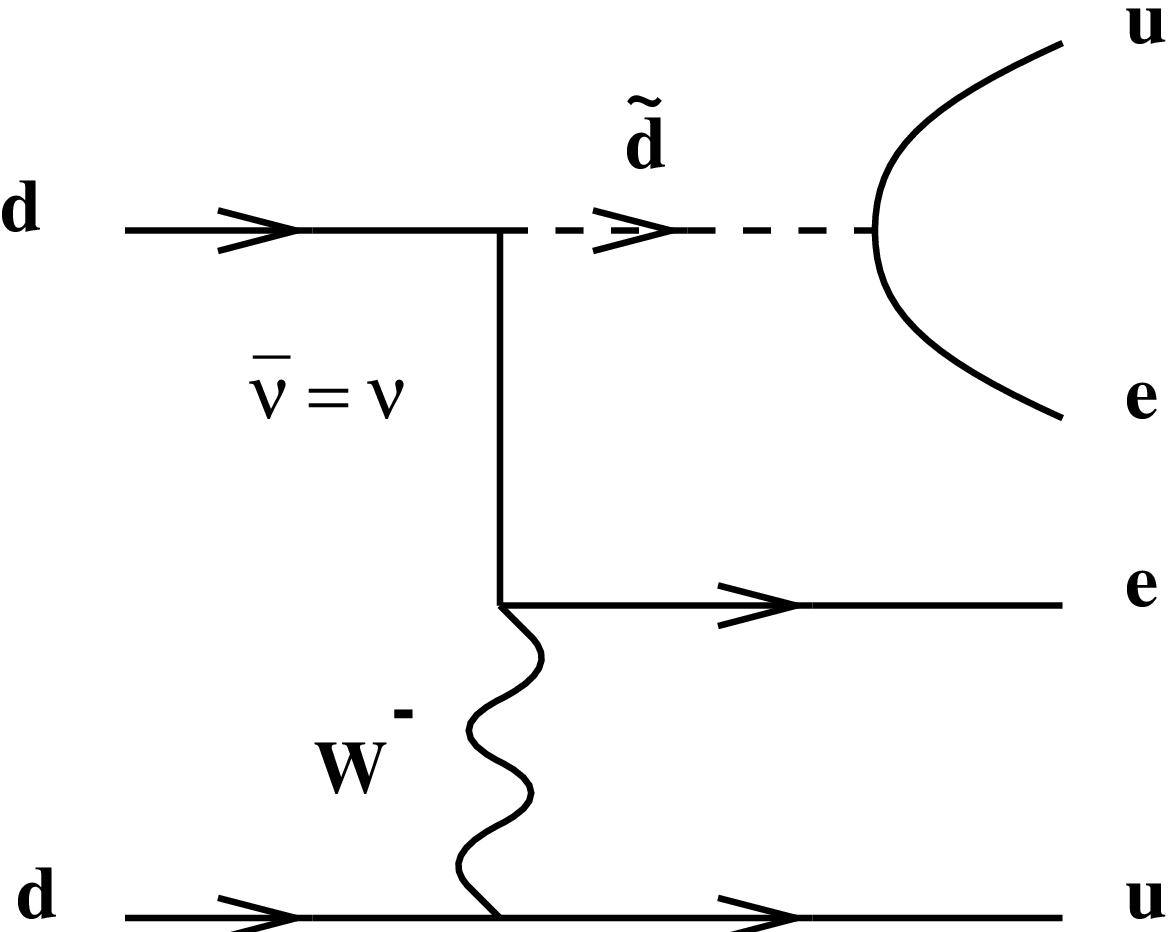}

\vskip-50mm
\hskip60mm
\epsfysize=50mm
\epsfxsize=50mm
\epsfbox{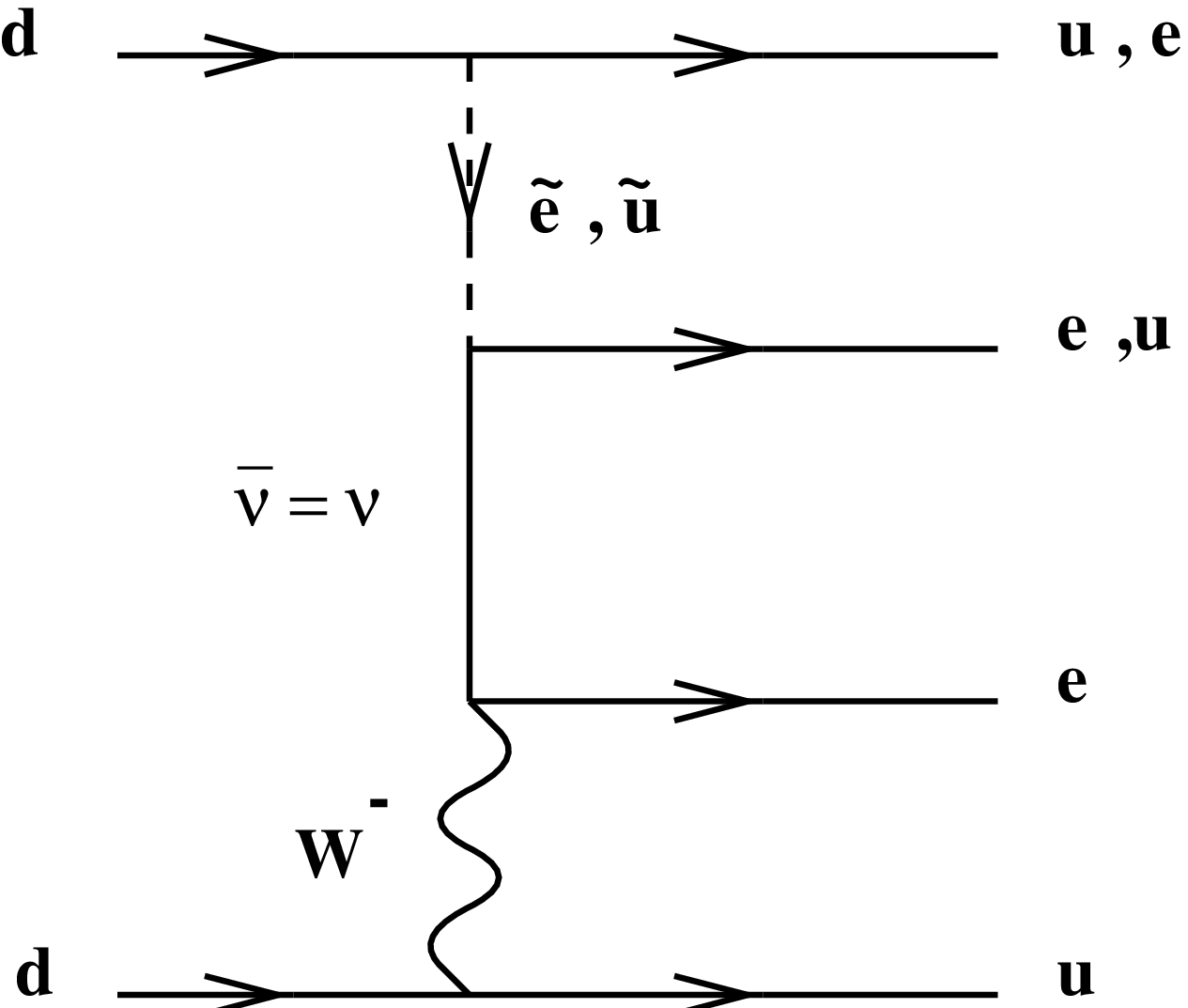}

\bigskip
\noindent
{\bf Figure 3: }{\it Leading double-beta-decay Feynman graphs in the
\rp model.}
\bigskip

\newpage
%
%
.
\vskip-30mm
%
%
\epsfysize=100mm
\epsfxsize=100mm
\hskip20mm
\epsfbox{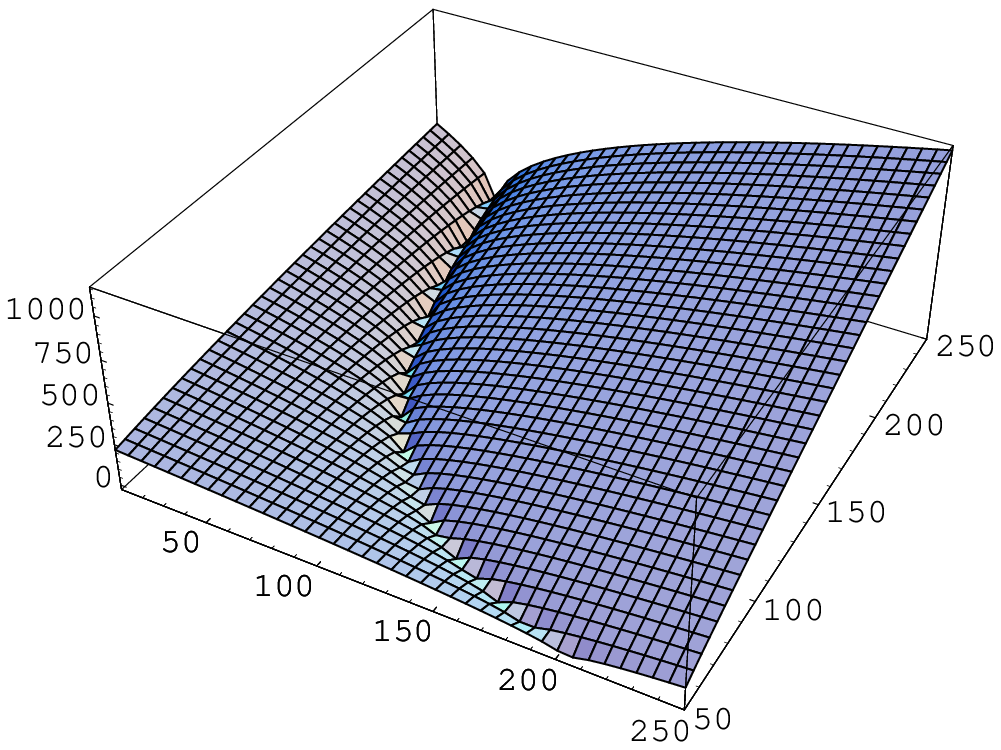}

\vskip-10mm
\hskip50mm
$M_2$ [GeV]

\vskip-50mm
\noindent
$\epsilon_1$  [keV]

\vskip20mm
\hskip110mm
$\mu$  [GeV]

\vskip20mm
%
%
\bigskip
\noindent
{\bf Figure 4.a: }{\it Upper limit on the individual parameter
$\epsilon_1$ for $\tan\beta=1$ and $\omega_i=0$ as a function of
($M_2,\mu$).}

\vskip-10mm
%
%
\epsfysize=100mm
\epsfxsize=100mm
\hskip20mm
\epsfbox{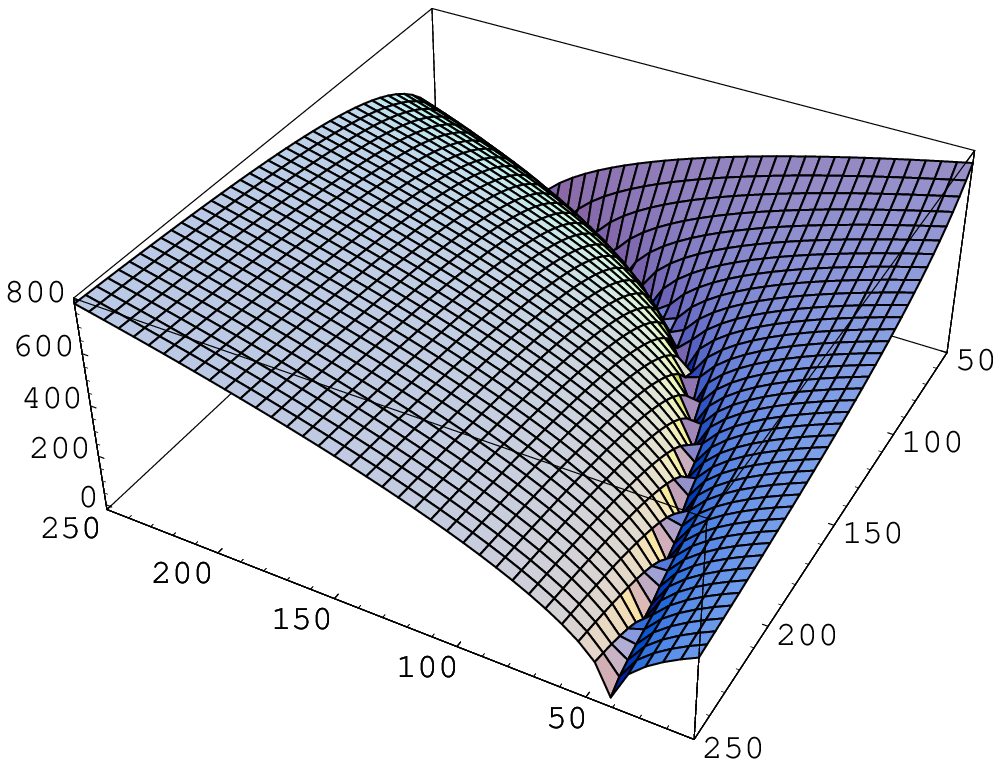}

\vskip-10mm
\hskip50mm
$M_2$ [GeV]

\vskip-50mm
\noindent
$\omega_1$  [keV]

\vskip20mm
\hskip110mm
$\mu$  [GeV]

\vskip20mm
%
%
\bigskip
\noindent
{\bf Figure 4.b: }{\it Same as figure 4.a, but for $\omega_1$, 
assuming $\epsilon_1=0$.}

\newpage

.
\vskip-30mm
%
%
\epsfysize=120mm
\epsfxsize=120mm
\hskip10mm
\epsfbox{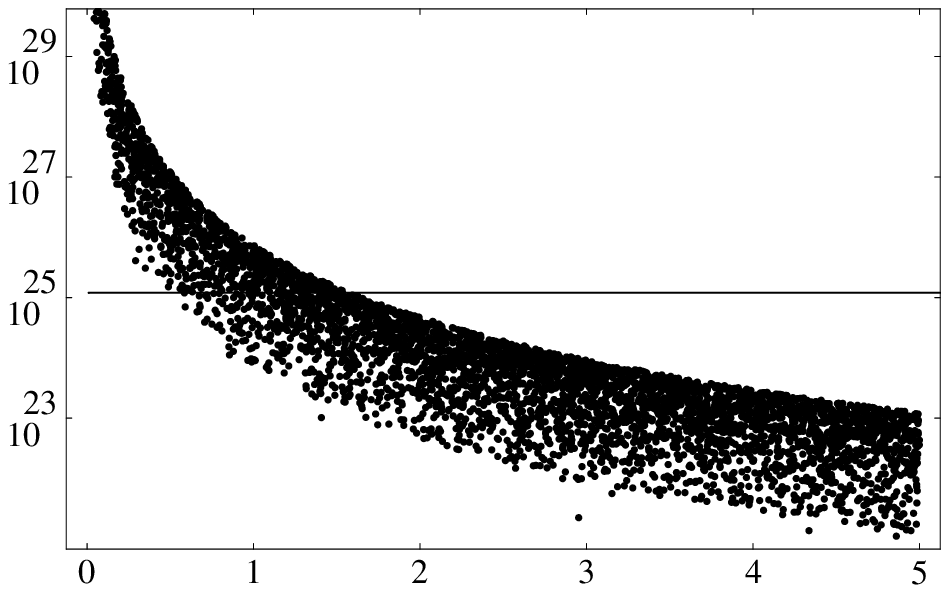}

\vskip-20mm
\hskip110mm
$\omega_1$  [MeV]

\vskip-70mm
\noindent
$T_{1/2}^{\znbb}$ [ys]

\vskip70mm
%
%
\bigskip
\noindent
{\bf Figure 5: }{\it Calculated half-life for the $\znbb$ decay 
of $^{76}Ge$ as function of $\omega_1$ for a random variation 
of the MSSM parameters, $M_2$ and $\mu$ from $100$ GeV to $1$ TeV 
and $\tan\beta=1-50$.}

\newpage

.
\vskip-30mm
%
%
\epsfysize=100mm
\epsfxsize=100mm
\hskip20mm
\epsfbox{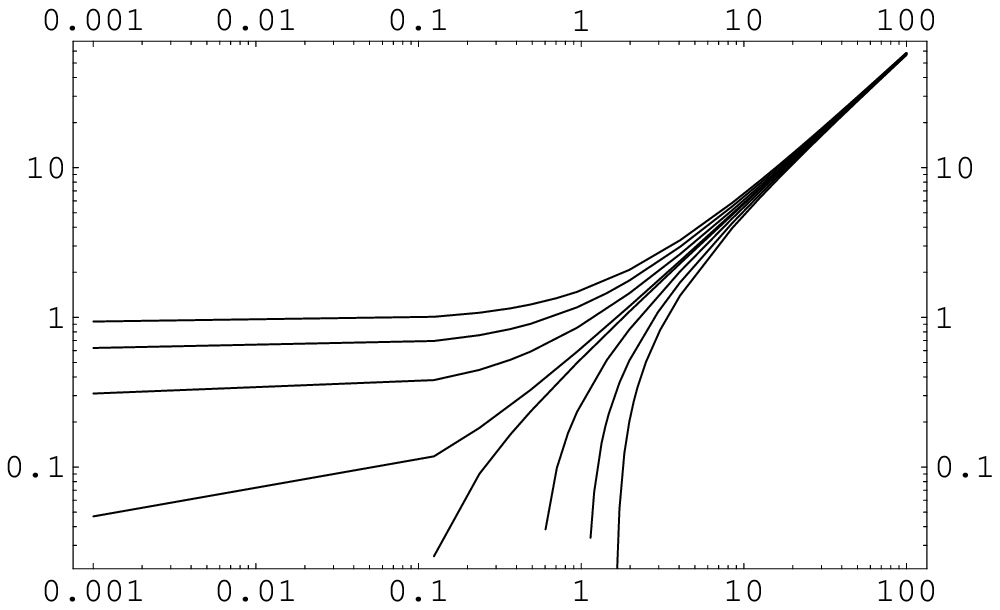}

\vskip-20mm
\hskip90mm
$\omega_1$ [MeV]

\vskip-50mm
\noindent
$\epsilon_1$  [MeV]

\vskip50mm
%
%
\bigskip
\noindent
{\bf Figure 6.a: }{\it Upper limits on ($\epsilon_1,\omega_1$)
assuming $\langle m_{\nu}\rangle \le 0.5$ eV, $\tan\beta=1$ and
$\mu=100$ GeV, for different values of $M_2$, $M_2= 100, 200, 500,
1000$ GeV.  Note that the allowed range is always in between two lines
of constant $M_2$.}

\vskip-10mm
%
%
\epsfysize=100mm
\epsfxsize=100mm
\hskip20mm
\epsfbox{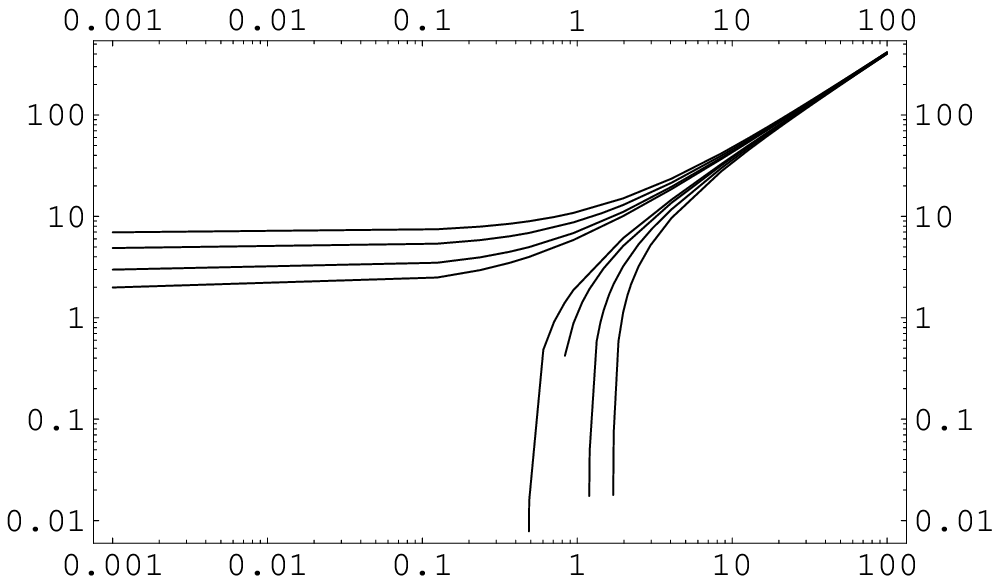}
\vskip-20mm
\hskip90mm
$\omega_1$ [MeV]

\vskip-50mm
\noindent
$\epsilon_1$  [MeV]

\vskip50mm
%
%
\bigskip
\noindent
{\bf Figure 6.b: }{\it Same as figure 6.a, but for $\tan\beta=10$.}

\end{document}